\documentclass[options]{kluwer}    
\newdisplay{guess}{Conjecture}

\usepackage[]{graphicx}

\begin{document}

\begin{article}
\begin{opening}         
\title{Do clusters of galaxies rotate?} 
\author{Hrant M. Tovmassian}  
\runningauthor{H. M Tovmassian}
\runningtitle{Clusters of galaxies}
\institute{Instituto Nacional de Astrof\'{\i}sica \'Optica y Electr\'onica,
AP 51 y 216, 72000, Puebla, Pue, M\'exico}
\date{September 2002}

\begin{abstract}
It is shown that the number of members, $N_A$, and the linear radius of the 
Abell clusters at $z<0.18$ increase with redshift. The increase is caused by 
observational selection: the number of rich and large clusters is small in the 
nearby small volume of space, while poor and small clusters are not detected 
at high distances. Besides, the numbers of clusters with different 
ellipticities are about the same at all distances, though in the case of 
cosmological evolution the number of the more elliptical clusters increases 
with decrease of distance. It means that clusters do not evolve dynamically. 
The dependences of the X-ray emission detection rate and the temperature of 
the X-ray emission on redshift are consequences of the dependence of $N_A$ on 
redshift. The radial velocity dispersion of the less elliptical clusters is 
higher than that of the elongated ones. This shows that galaxies in a cluster 
move preferentially along the large axis of the cluster. The space density is 
not changing in the considered range of $z$. Movement of galaxies along the 
cluster elongation may be explained by rotation along the large axis of a 
cluster.

\end{abstract}

\end{opening}

It is supposed that clusters grow by accretion and merging of galaxies along 
filamentary large-scale structures (West 1994, West et al. 1995). Plionis 
(2002) claims that nearby clusters at $z<0.15$ are still evolving. It means, 
that nearby clusters should be richer and perhaps larger than the distant 
ones. I compared the number $N_A$ of galaxies (Struble \& Rood 1987) in ACO 
clusters (Abell et al. 1989) and their linear radii $R_l$ (in Mpc) with 
redshift $z$ (Figs. 1 and 2). For analysis I used 621 clusters with $z$ 
determined by at least 2 galaxies (Struble \& Rood 1999) and located at 
$z<0.18$. Figs. 1 and 2 show that $N_A$ and $R_l$ increase with $z$. The 
dashed regression lines in both Figures are drawn after exclusion of 381 
clusters of the richness class 0, the sample of which is assumed to be 
incomplete. The Spearman rank test showed that the correlation factors for the 
total sample and for the sample without richness class 0 clusters are 
$\rho=0.41$ and $\rho=0.42$ with significances $\alpha=10^{-27}$ and 
$\alpha=10^{-12}$ respectively. Corresponding values for $R_l$ (Fig. 2) are 
$\rho=0.44$ and $\rho=0.37$, and $\alpha=10^{-31}$ and $\alpha=10^{-9}$ 
respectively. Since exclusion of the richness class 0 clusters does not alter 
these dependences, in the further analysis I use the whole sample of clusters. 
Apparently the dependences of $N_A$ and $R_l$ on $z$ {\it are due to a 
selection effect}. At high $z$ mostly rich and large clusters are detected. 
The number of them is small in the nearby small volume of space. Poor and 
small clusters located at high $z$ are missed in the background. If 
dependences of $N_A$ and $R_l$ on $z$ are caused by evolution, $N_A$ and $R_l$ 
should be correlated. However, comparison shows no correlation ($\rho=0.119$, 
$\alpha=0.003$) between $N_A$ and $R_l$. 

\begin{figure}[htb]
\vspace{0.35cm}
\includegraphics[width=14pc]{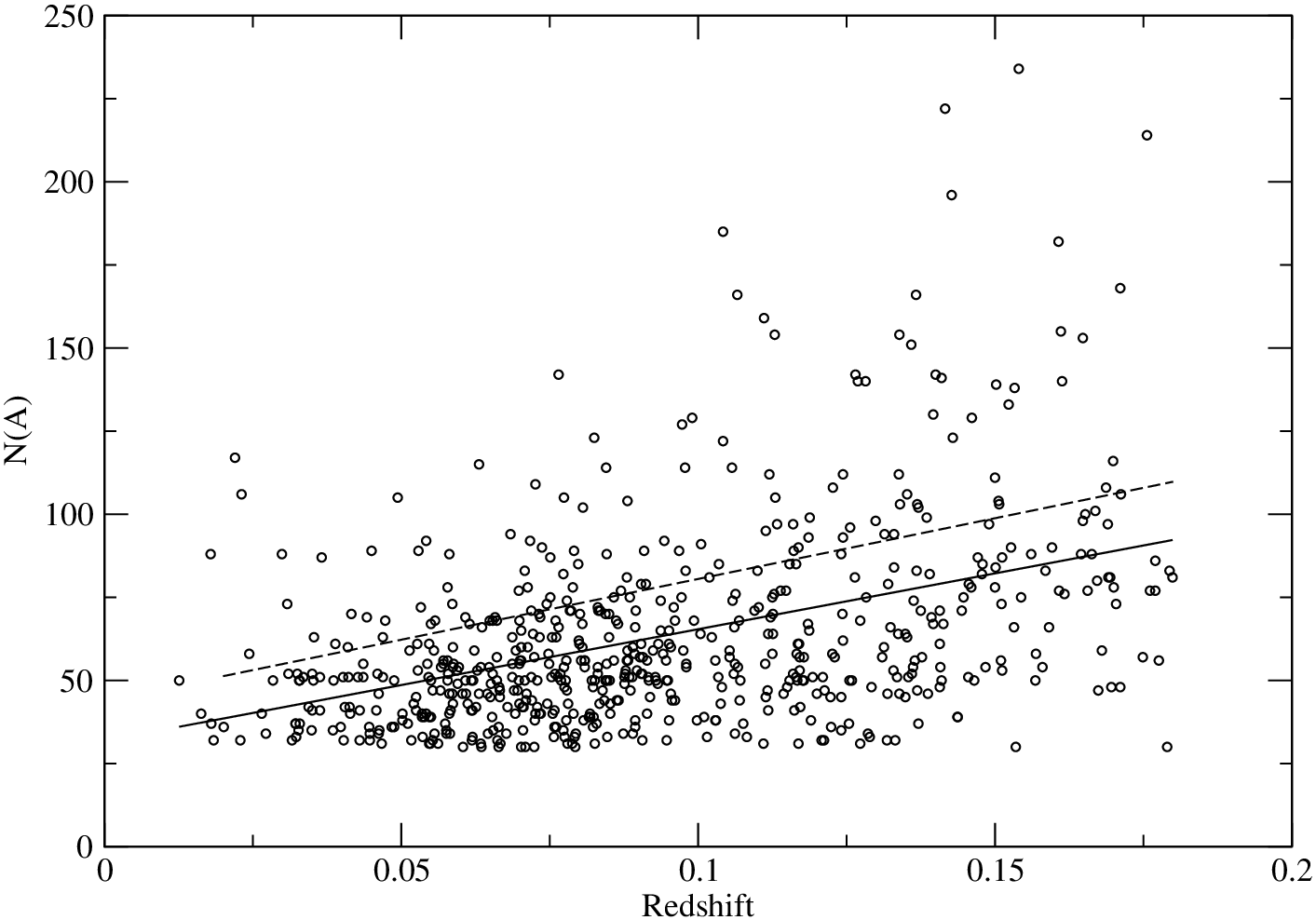}
\caption{The dependence of $N_A$ on $z$. The solid line is the 
regression line determined using the data on 616 clusters. The dashed line is 
determined for a sample in which 378 poor clusters of richness 0 are 
excluded.}
\label{fig. 1}
\end{figure}

\begin{figure}[htb] 
\centerline{\includegraphics[width=14pc]{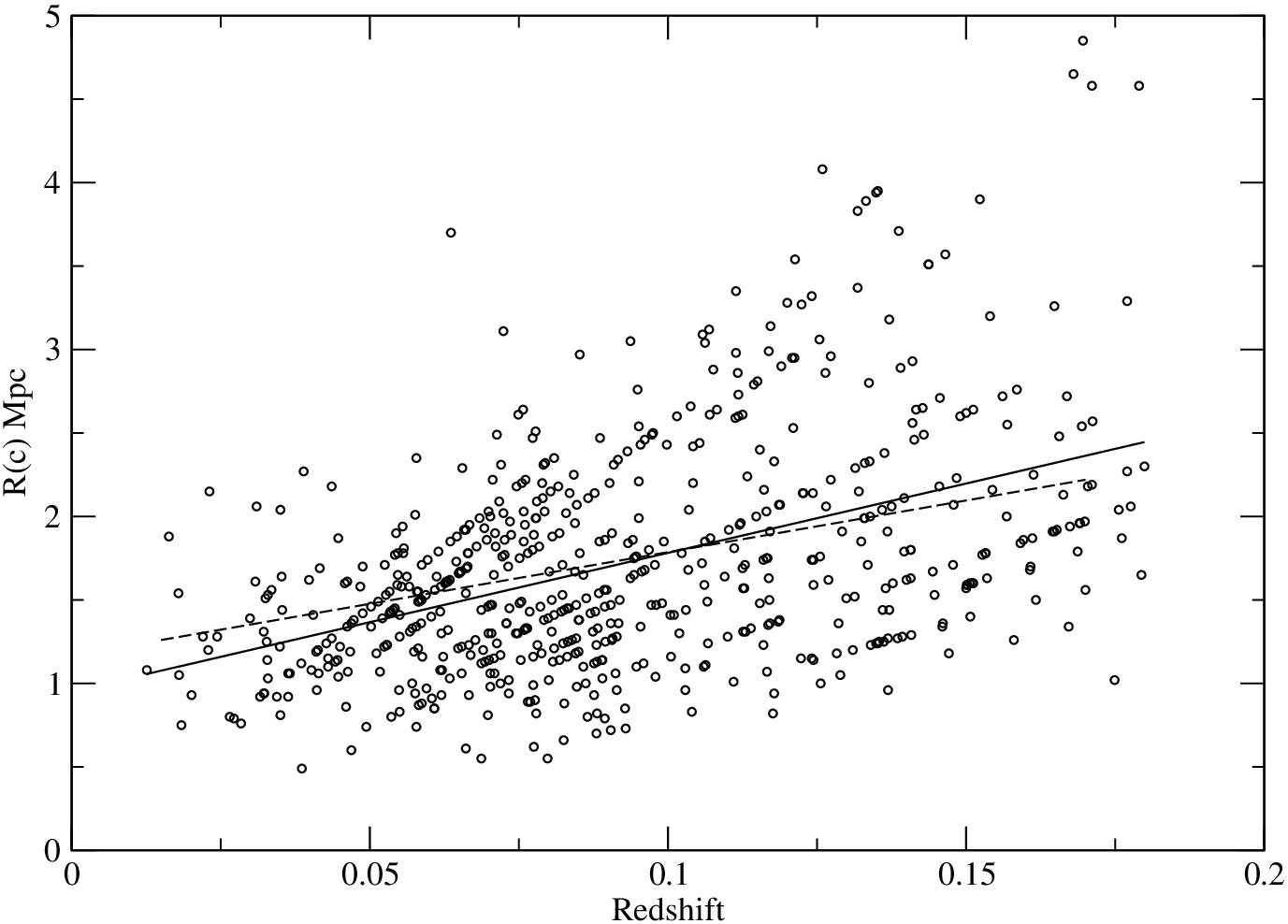}}
\caption[]{The dependence of the linear $R_c$ on $z$. The solid and dashed 
lines are determined as in Fig. 1.}
\label{fig. 2}
\end{figure}

Besides, in the case of cosmological evolution the number of elliptical 
clusters (ECs) should be higher at high $z$, and the number of round clusters 
(RCs) should be high at small $z$ (Mellot et al. 2001). I compared the 
distribution of the number of ECs with $f\leq1.18$\footnote{$f=a/b$ where $a$ 
is the length of the cluster and $b$ is its width.} with that of the RCs with 
$f\geq2.0$ (Fig. 3). The values of $f$ are taken from Struble \& Ftaclas 
(1995). The K-S test showed that the hypothesis that two distributions are of 
different population is rejected with probability 0.6, i.e. both samples are 
consistent with a single distribution function. Moreover, the RCs should 
accumulate in time more members and should be more rich than ECs. Meanwhile, 
the mean number of members of 18 RCs is $61\pm22$, and does not differ from 
that of the 27 ECs, being equal to $63\pm32$. In the case of cosmological 
evolution the cluster velocity dispersion, $\sigma_v$, should increase in time 
since virialization will tend to increase the cluster ``thermal'' velocity 
dispersion (e.g. Plionis 2002). I considered $\sigma_v$'s of a sample of 362 
clusters with $\sigma_v$ determined by at least five galaxies (Struble \& Rood 
1999). The Spearman rank test showed that, contrary to expectation, $\sigma_v$ 
is not correlated with $z$ ($\rho=0.128$ and the significance 
$\alpha=0.0146$).  

\begin{figure}[htb] 
\vspace{0.2cm}
\centerline{\includegraphics[width=10pc]{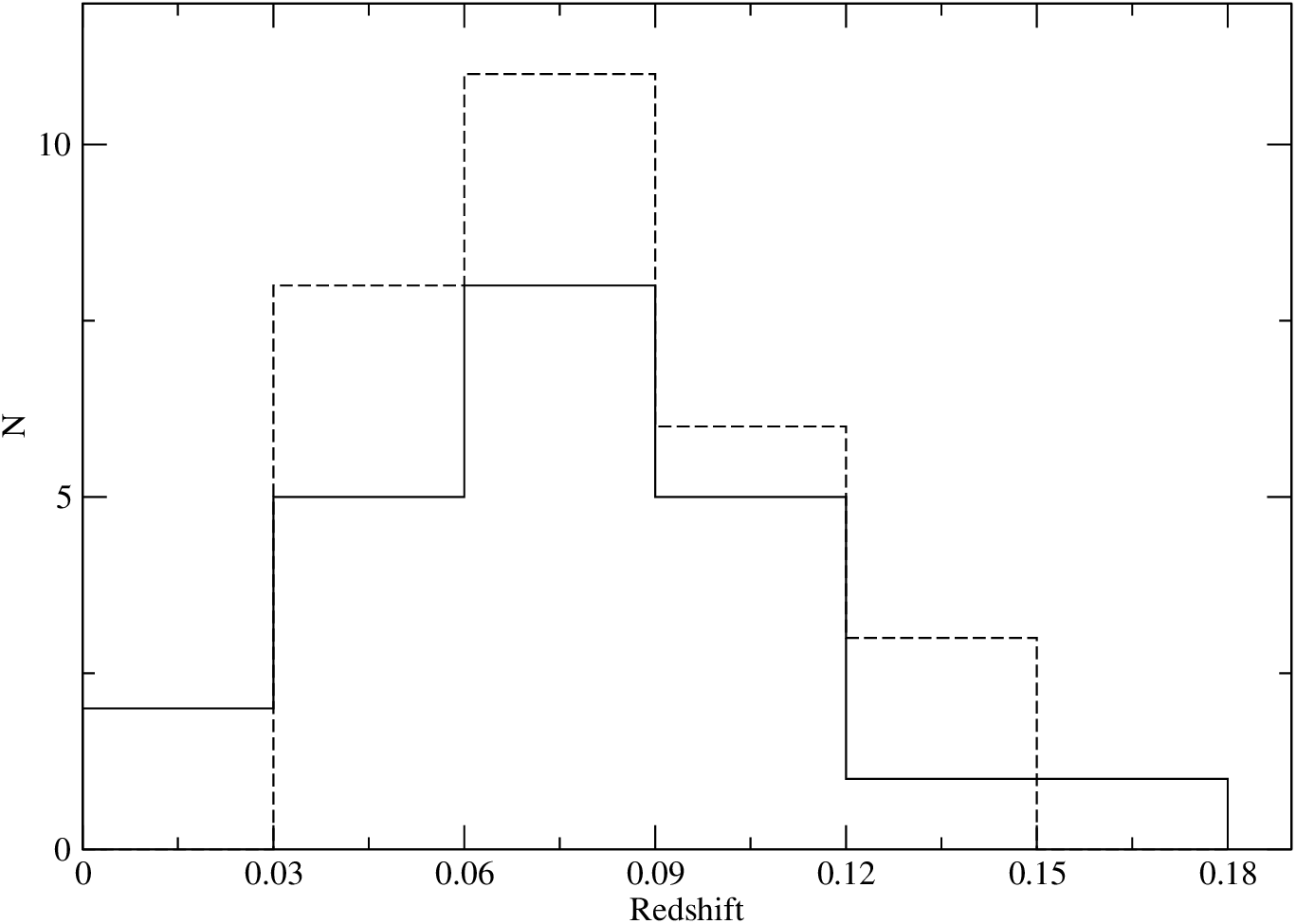}}
\caption[]{The distribution of the numbers $N$ of the RCs (solid line) and ECs 
(dashed line) by $z$.}
\label{fig. 3}
\end{figure}

If clusters grow by accretion of the surrounding matter, then the space 
density $\rho$ of galaxies should increase in time, i.e. it should be higher 
in the nearby space. Fig. 4 shows that $\rho$ increases by 4.5 times from 
$z=0.18$ to $z=0.02$. However, just by this amount $\rho$ must increase due to 
increse of $N_A$ and $R_l$ with $z$ (Figs. 1 and 2) which, in turn, is a 
result of observational selection.

\begin{figure} 
\centerline{\includegraphics[width=14pc]{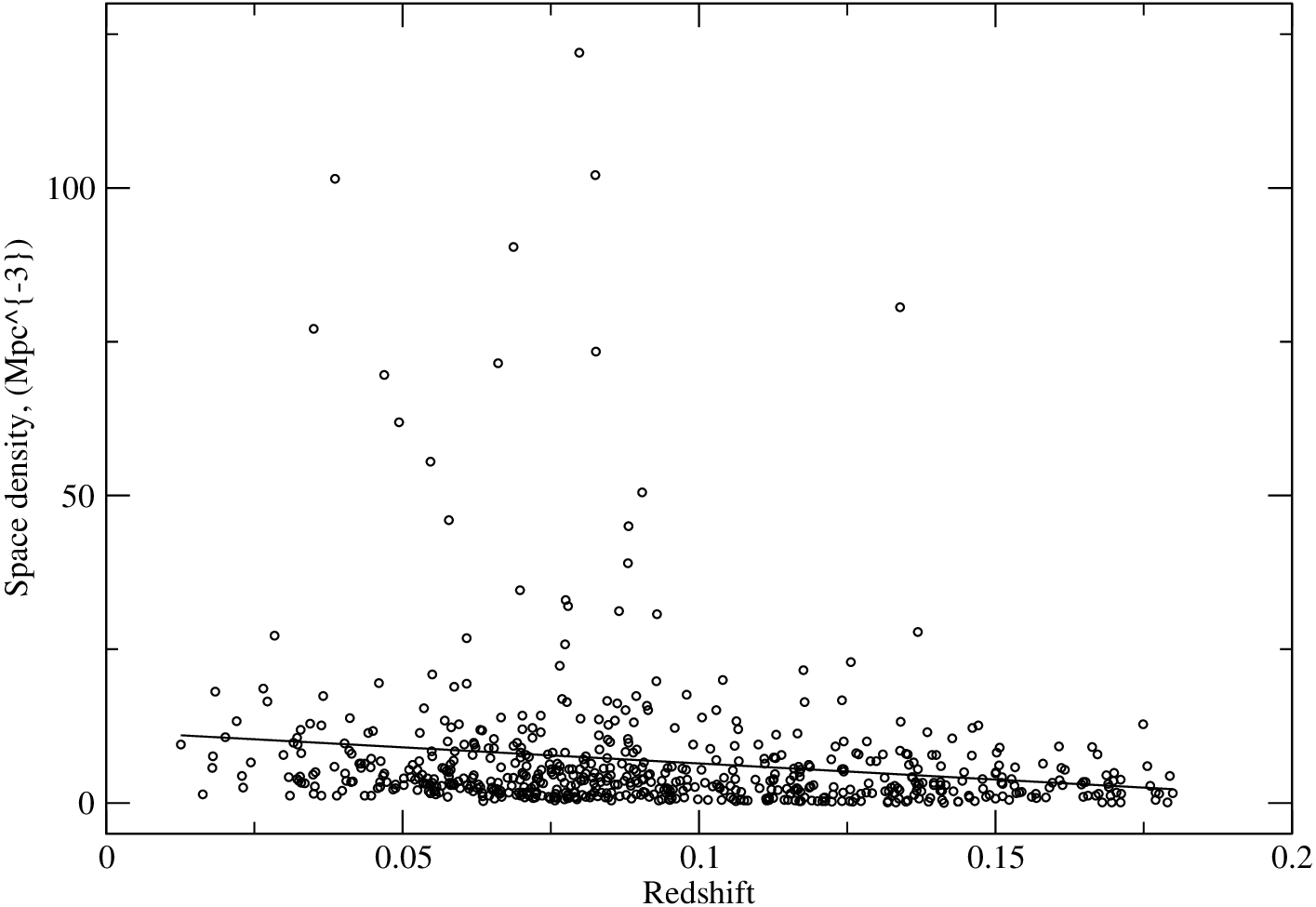}}
\caption[]{The dependence of the space density of galaxies in clusters on $z$.}
\label{fig. 4}
\end{figure}

Hence, observational data {\it contradict to cosmological evolution} of 
clusters up to $z=0.18$.  

The observed dependence of $N_A$ on $z$ may cause correlation between other 
cluster parameters. In Table 1 the average $N_A$ of clusters with detected 
X-ray emission is compared with that of clusters with no X-ray detection 
(Ebeling et al. 1996, 1998, 2000). It is seen that in each of the four ranges 
of $z$ the average $N_A$ of clusters with X-ray emission is higher than in 
clusters without X-ray emission. In Fig. 5 the ICM gas temperature $T$ of 213 
clusters is compared with $N_A$. It shows that $T$ increases with $N_A$: 
$log~kT \sim 0.54 log~N_A$. The correlation factor $\rho=0.54$ and 
significance $\alpha=1.7\cdot10^{-17}$, i.e the correlation is strong. On the 
other hand, it has been shown that the temperature $T$ of the X-ray emitting 
gas and its mass $M_{gas}$ are correlated: $M_{gas} \sim T^{1.5}$ (Evrard et 
al. 1996) or $M_{gas} \sim T^{1.79}$ (Nevalainen et al. 2000) at $z=0$. It 
follows that $M_{gas} \sim N_A^{0.96}$, i.e. the ISM mass is strictly 
proportional to the number of galaxies. Thus, the correlations of the X-ray 
detection rate and the X-ray temperature $T$ on $z$ are simply consequences of 
the dependence of $N_A$ on $z$, caused by observational selection.

{\scriptsize
\begin{table}
\caption[]{The average numbers $N_A$ of clusters with and without X-ray 
emission.}
\begin{tabular}{ccccc}
\hline
 $z$   & $<0.06$ & 0.06-0.10 & 0.10-0.14 & 0.14-0.18 \\
\hline
D-ted & $54.7\pm19.4$ (57) & $63.4\pm21.7$ (65) & $90.8\pm33.2$ (23) &
$131.2\pm64.7$ (10) \\
Not d-ted & $46.5\pm14.2$ (69) & $54.1\pm25.5$ (191) & $64.3\pm27.7$ (128) &
$84.7\pm36.0$ (70) \\
\hline
\end{tabular}
\end{table}
}

\begin{figure} 
\centerline{\includegraphics[width=14pc]{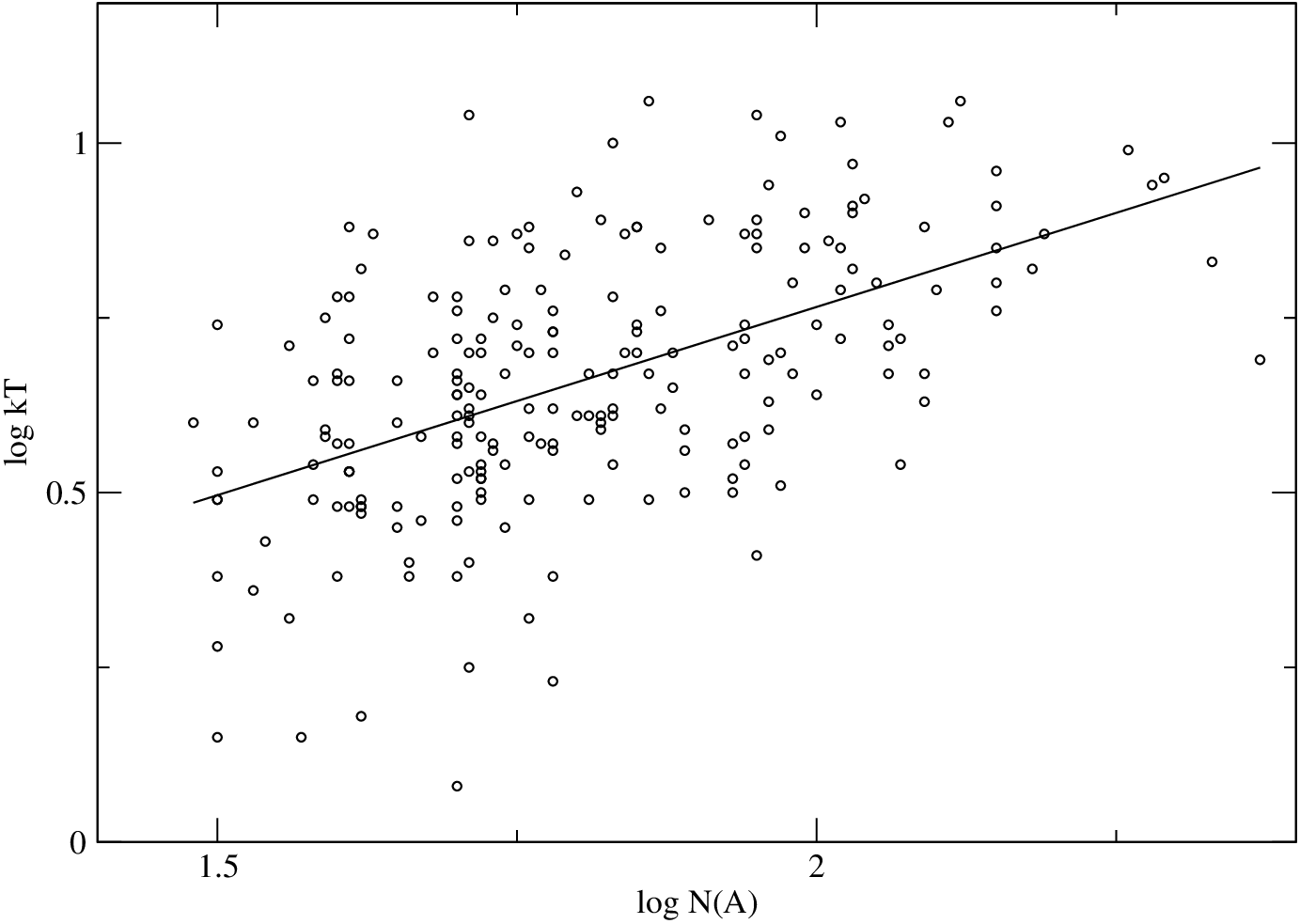}}
\caption[]{The dependence of kT on $N_A$.}
\label{fig. 5}
\end{figure}

Tovmassian (2001, 2002) showed that the radial velocity dispersion $\sigma_v$ 
of compact groups (CGs) is minimal for elongated chain-like groups with small 
$b/a$ ratio, while the less elongated groups have higher $\sigma_v$. It shows 
that the member galaxies in a group move preferentially along the elongation 
of a group. The correlation $\sigma_v$ versus $b/a$ holds in the case of 
clusters as well. The mean $\sigma_v$ of 15 RCs is $917\pm338$, and that of 18 
ECs is $642\pm254$. The $t$ test shows that the difference is highly 
significant ($\alpha=0.012$). The mean $N_A$ of 12 RCs is $62\pm17$, while 
that of 17 ECs is $60\pm25$. It means that the difference in $\sigma_v$ is not 
caused by the difference in $N_A$. Therefore, we conclude, that the observed 
difference of $\sigma_v$ in RCs and ECs is due to a {\it projection} effect. 
Hence, galaxies in clusters, as in CGs, move {\it preferentially along the 
large axis}. In clusters the large axis of which is oriented close to the line 
of sight and which are seen more round, a large number of member galaxies move 
towards us or backwards. In such clusters $\sigma_v$ is high, and is close to 
the real value. The $\sigma_v$ of ECs oriented, on average, at $\sim45\deg$ to 
the line of sight should be by $\approx sin~45\deg$ smaller than those of RCs, 
i.e. should be equal to 648 km s$^{-1}$. The mean $\sigma_v$ of ECs differs 
only by $\approx 10\%$ from the estimated value. 

Movement of galaxies along the large axis of a cluster may be explained by 
three ways. The option that clusters grow by accretion and merging of matter 
mainly along the large axis (West 1994, West et al. 1995) is hardly valid, 
since as we showed, observational data contradict to this option. Moreover, in 
the case of accretion only a few infalling galaxies would move along the 
cluster elongation, but not the majority of its members. Another option is 
departure of member galaxies from each other in opposite directions. The 
reason may probably be the influence of another clusters. However, the 
velocities initiated by the gravity of clusters with masses $\sim10^{14}$ 
located at opposite directions at distances of $\approx30$ Mpc are $\approx 
30$ km s$^{-1}$, and may not account for the observed difference of 
$\sigma_v$. No physical mechanism is yet known for explanation of expansion of 
clusters according to Ambartsumian (1961) or Arp (2001, and references 
therein) ideas. Hence, the more realistic option is {\it rotation of galaxies} 
around their gravitational center, as in the case of CGs (Tovmassian 2001, 
2002).

\end{article}

\begin{thebibliography} {}

\bibitem[1989]{aco} Abell, G.O., Corwin, H.C., Olowin, R., 1989, ApJS,
70, 1

\bibitem[1961]{amb} Ambartsumian V.A. 1961, AJ, 66, 536    

\bibitem[1999]{arp99} Arp H. 2001, ApJ, 549, 780

\bibitem[1996]{eb96} Ebeling, H., Voges, W., B\"{o}hringer, H., et al., 1996, 
MNRAS, 281, 799

\bibitem[1998]{eb98} Ebeling, H., Edge, A. C., B\"{o}hringer, H., 1998, MNRAS,
301, 881

\bibitem[2000]{eb00} Ebeling, H., Edge, A.C., Allen, S.W., et al., 2000, 
MNRAS, 318, 333

\bibitem[2001]{mcm} Melott, A.L., Chambers, S.W., Miller, C.J. 2001, ApJ, L75

\bibitem[2000]{nmf} Nevalainen, J., Markevitch, M., Forman, W. 2000, ApJ, 
532, 694

\bibitem[2002]{pl} Plionis, M. 2002, ApJ, 572, 67L

\bibitem[1995]{sf} Struble, M. F., Ftaclas, C., 1995, AJ, 108, 1

\bibitem[1987]{sr87} Struble, M. F., Rood, H. J., 1987, ApJS, 63, 555

\bibitem[1999]{sr99} Struble, M. F., Rood, H. J., 1999, ApJS, 125, 35


\bibitem[2001]{t01} Tovmassian, H. M., 2001, PASP, 113, 543

\bibitem[2000]{t02} Tovmassian H. M.,  2002, AN, 323, 488  

\bibitem[1994]{west} West, M. J. 1994, MNRAS, 268, 79

\bibitem[1995]{wjf} West, M.J., Jones, C., Forman, W. 1995, ApJ, 451, L5

\end{thebibliography}
\end{document}